\documentclass[twocolumn]{aastex631}

\hypersetup{linkcolor=blue, citecolor=blue, filecolor=cyan, urlcolor=blue}

\usepackage{multirow}
\usepackage{amsmath}

\shorttitle{Merger in LAEs}
\shortauthors{J. Ren et al.}
\graphicspath{{./}{figures/}}
\begin{document}

\correspondingauthor{F. S. Liu and Nan Li}
\email{E-mail: fsliu@nao.cas.cn; nan.li@nao.cas.cn}

 \title{The JWST Unveils the Bimodal Nature of Lyman Alpha Emitters at $3< z<7$: Pristine versus Merger-Driven Populations}

\author[0000-0002-5043-2886] {Jian Ren}
\affil{National Astronomical Observatories, Chinese Academy of Sciences, 20A Datun Road, Chaoyang District, Beijing 100101, China}
\affil{Key Laboratory of Space Astronomy and Technology, National Astronomical Observatories, Chinese Academy of Sciences, 20A Datun Road, Chaoyang District, Beijing 100101, China}

\author[0009-0001-7105-2284]{F. S. Liu $^{\color{blue} \dagger}$}
\affil{National Astronomical Observatories, Chinese Academy of Sciences, 20A Datun Road, Chaoyang District, Beijing 100101, China}
\affil{Key Laboratory of Optical Astronomy, National Astronomical Observatories, Chinese Academy of Sciences, 20A Datun Road, Chaoyang District, Beijing 100101, China}
\affil{School of Astronomy and Space Science, University of Chinese Academy of Science, Beĳing 100049, China}

\author{Nan Li $^{\color{blue} \dagger}$}
\affil{National Astronomical Observatories, Chinese Academy of Sciences, 20A Datun Road, Chaoyang District, Beijing 100101, China}
\affil{Key Laboratory of Space Astronomy and Technology, National Astronomical Observatories, Chinese Academy of Sciences, 20A Datun Road, Chaoyang District, Beijing 100101, China}
\affil{School of Astronomy and Space Science, University of Chinese Academy of Science, Beĳing 100049, China}

\author{Qi Song}
\affil{National Astronomical Observatories, Chinese Academy of Sciences, 20A Datun Road, Chaoyang District, Beijing 100101, China}
\affil{Key Laboratory of Optical Astronomy, National Astronomical Observatories, Chinese Academy of Sciences, 20A Datun Road, Chaoyang District, Beijing 100101, China}

\author{Pinsong Zhao}
\affil{Kavli Institute for Astronomy and Astrophysics, Peking University, Beijing 100871, China}
\affil{National Astronomical Observatories, Chinese Academy of Sciences, 20A Datun Road, Chaoyang District, Beijing 100101, China}

\author[0009-0001-5320-1450]{Qifan Cui}
\affil{Shanghai Key Lab for Astrophysics, Shanghai Normal University, Shanghai 200234, China}
\affil{National Astronomical Observatories, Chinese Academy of Sciences, 20A Datun Road, Chaoyang District, Beijing 100101, China}

\author{Yubin Li}
\affil{National Astronomical Observatories, Chinese Academy of Sciences, 20A Datun Road, Chaoyang District, Beijing 100101, China}
\affil{Key Laboratory of Space Astronomy and Technology, National Astronomical Observatories, Chinese Academy of Sciences, 20A Datun Road, Chaoyang District, Beijing 100101, China}

\author{Hao Mo}
\affil{National Astronomical Observatories, Chinese Academy of Sciences, 20A Datun Road, Chaoyang District, Beijing 100101, China}
\affil{Key Laboratory of Optical Astronomy, National Astronomical Observatories, Chinese Academy of Sciences, 20A Datun Road, Chaoyang District, Beijing 100101, China}
\affil{School of Astronomy and Space Science, University of Chinese Academy of Science, Beĳing 100049, China}

\author{Guanghuan Wang}
\affil{Purple Mountain Observatory, Chinese Academy of Sciences, 10 Yuanhua Road, Nanjing 210034, China}
\affil{National Astronomical Observatories, Chinese Academy of Sciences, 20A Datun Road, Chaoyang District, Beijing 100101, China}

\author{Hassen M. Yesuf}
\affil{Key Laboratory for Research in Galaxies and Cosmology, Shanghai Astronomical Observatory, Chinese Academy of Sciences, 80 Nandan Road, Shanghai 200030, China}

\author{Weichen Wang}
\affil{Dipartimento di Fisica G. Occhialini, Università degli Studi di Milano-Bicocca, Piazza della Scienza 3, I-20126 Milano, Italy}

\begin{abstract}

We present a systematic study of merging galaxies among Lyman-alpha emitters (LAEs) using JWST/NIRCam high-resolution 
imaging data. From a large sample of 817 spectroscopically confirmed LAEs at $3<z<7$ in the GOODS-S field, we identify late-stage mergers and interacting systems with fractions of $39.4\%\pm2.5\%$ and $60.6\%\pm6.3\%$, respectively. These fractions exhibit significant redshift evolution and depend on both stellar mass ($M_*$) and UV magnitude ($M_{\rm UV}$), 
being most prevalent in massive ($\log(M_*/M_\odot)>8.5$) and bright ($M_{\rm UV}<-19.5$) systems. 
At fixed $M_*$ and $M_{\rm UV}$, we find negligible differences in the UV slope ($\beta$) between late-stage mergers and isolated LAEs; however, a clear bimodal distribution emerges in the $M_*$-sSFR plane, 
where isolated LAEs peak at $\log(M_*/M_\odot)\approx7.8$ and $\log({\rm sSFR/yr^{-1}})\approx-7.4$, 
and late-stage mergers peak at $\log(M_*/M_\odot)\approx8.6$ and $\log({\rm sSFR/yr^{-1}})\approx-7.6$. 
Our results reveal two evolutionary classes---Pristine LAEs, low-mass ($M_*<10^{8.5}M_\odot$), isolated systems that represent early-stage galaxies with minimal merger interactions, and Merger-driven LAEs, massive ($M_*>10^{8.5}M_\odot$) systems in which mergers enhance star formation and facilitate the escape of Lyman-alpha photons or accrete pristine LAEs---both of which are consistent with both observational and theoretical expectations and collectively demonstrate that mergers are a central driver of LAE evolution across the first two billion years.
 
\end{abstract}

\keywords{Galaxy morphology---Merger---Lyman Alpha emission line Galaxies}

\section{Introduction} \label{sec:intro}

Lyman-alpha emitters (LAEs) are key tracers of the early universe, offering valuable insights into the epoch of reionization and the formation of primordial galaxies \citep{Partridge1967, Dijkstra2014, Ouchi2018}. The resonant Ly$\alpha$ emission line is highly sensitive to the neutral hydrogen content of the intergalactic medium (IGM) and the escape fraction of ionizing photons, making LAEs powerful probes for studying cosmic reionization \citep{Inoue2014}. Furthermore, these objects serve as unique laboratories for exploring the interplay between galaxy morphology, star formation, and feedback processes--key factors that drive galaxy evolution \citep{Dijkstra2014, Dayal2014}

Previous studies have characterized high-redshift ($z>3$) LAEs as typically young, compact, low-mass ($\log(M_*/M_\odot) \lesssim 8$) systems with little dust, and they are often considered potential progenitors of present-day Milky Way-mass 
galaxies \citep{Gawiser2007, Rauch2008, Ono2010, Gronwall2011, Guaita2011}. 
While the majority of LAEs exhibit compact morphologies \citep{Venemans2005, Pirzkal2007, Malhotra2012, Paulino2018}, 
a significant subset shows irregular structures, including multiple components and clumpy distributions 
\citep{Pirzkal2007, Bond2009, Ouchi2013, Gronwall2011, Sobral2015, Paulino2018}. 
Observations from the Hubble Space Telescope (HST) indicate merger fractions of $\sim$20\% 
at $z>2$ \citep{Bond2009, Cowie2010, Chonis2013, Shibuya2014, Kobayashi2016}, 
with brighter systems ($M_{\rm UV} < -21$) potentially exceeding 50\% \citep{Jiang2013}. 
These findings are consistent with simulations that predict major merger fractions exceeding $30\%$ at $z>3$, 
underscoring the importance of mergers in driving stellar mass assembly and star formation in LAEs \citep{Tilvi2011}.

Recent observations from the James Webb Space Telescope (JWST) have provided compelling evidence linking merger activity to the escape of Lyman-alpha (Ly$\alpha$) and Lyman-continuum (LyC) photons in high-redshift ($z>3$) galaxies. \citet{Witten2024} find ubiquitous close companions among $z>7$ LAEs, suggesting that mergers drive intrinsic Ly$\alpha$ emission and facilitate its escape. \citet{Yuan2024} detect spatially offset LyC emission in a merging LAE pair at $z=3.797$, associating the ionizing radiation with the younger, more actively star-forming component.  Similarly, \citet{Zhu2024} identify merging signatures in 20 out of 23 LyC leakers at $z=3-4.5$, indicating that mergers may enhance the escape of ionizing photons. 
However, other studies report more moderate merger fractions of $\sim$20\% at similar redshifts \citep{Liu2024, Ning2024}. 
Notably, \citet{Mascia2025} suggest that mergers may play a limited role in LyC escape for low-mass galaxies  ($\log(M_*/M_\odot) < 8$) at $z\leq5$, where compact morphology and intense star formation appear to be the dominant regulating factors.

Leveraging high-resolution imaging data from the JWST/NIRCam, this study presents a systematic analysis of 819 spectroscopically confirmed Lyman-alpha emitters (LAEs) within the redshift range $3<z<7$, aiming to (1) quantitatively characterize the evolution of merger fractions as functions of redshift, stellar mass ($M_*$), and UV magnitude ($M_{\rm UV}$), and (2) 
identify the physical differences between merging and isolated LAEs to uncover their underlying evolutionary pathways.

The paper is structured as follows: Section~\ref{sec:data} describes sample selection and data processing. Physical property measurements are detailed in Section~\ref{sec:methods}, with results presented in Section~\ref{sec:results}. We discuss implications for LAE evolution in Section~\ref{sec:discussion} and summarize conclusions in Section~\ref{sec:conclusions}. We adopt a $\Lambda$CDM cosmology with $H_0 = 70~\mathrm{km~s^{-1}~Mpc^{-1}}$, $\Omega_{\rm m} = 0.3$, $\Omega_{\Lambda} = 0.7$. UV magnitudes correspond to the GALEX FUV (1500\,\AA) filter bandpass.

\section{Data and sample } \label{sec:data}

\subsection{JWST imaging data}

We acquired raw JWST/NIRCam imaging data for the GOODS-S field spanning filters F090W, F115W, F150W, F200W, F277W, F356W, F444W, and medium bands (F182M, F210M, F335M, F410M) through Programs 1176, 1180, 1210, 1283, 1286, 1895, 1963, 2079, 2198, 2514, 2516, 3215, 3990, and 6541. 

Data reduction commenced with Stage 1 processing of uncalibrated data using standard pipeline parameters for initial detector corrections. Following mitigation of "snowball" artifacts, Stage 2 processing incorporated subtraction of "wisp" artifacts and $1/f$ noise reduction via the methodology of \citet{Schlawin2020}. Custom masks were applied to address persistence, dragon breath, ginkgo leaf, wisp features, and other systematic artifacts. For Stage 3, we implemented a custom astrometric solution: initial WCS calibration utilized HST/F160W imaging from CANDELS \citep{Grogin2011, Koekemoer2011} as reference for JWST/F150W data, with subsequent bands aligned to this reference frame. We then executed the \texttt{OutlierDetection} step followed by tailored source masking and background subtraction prior to resampling. Final mosaics were constructed per observational epoch, with localized background subtraction applied to sub-regions for optimal sky correction.

The reduced NIRCam dataset (v5) covers 180.54~arcmin$^2$ in GOODS-S with uniform 0.03$''$ pixel scale across all bands. The point spread function (PSF) exhibits a full width at half maximum (FWHM) of 0.164$''$ at F444W. 
All reduced imaging products will be publicly accessible via the JWST-SPRING 
archive\footnote{\url{http://groups.bao.ac.cn/jwst_spring/}} and described comprehensively in Liu et al. (2025, in preparation).

\subsection{LAE Data}

We construct a spectroscopically confirmed sample of Ly$\alpha$ emitters (LAEs) at $3 \leq z \leq 7$ in GOODS-S by combining detections from three major surveys: MUSE \citep{Herenz2017,Bacon2017}, VANDELS \citep{McLure2018}, and CANDELS-z7 \citep{Pentericci2018}. 

The MUSE component incorporates data from two programs: (1) The MUSE-Wide survey \citep{Herenz2017,Urrutia2019} covers $4750-9350$~\AA\ with 1-hour exposures, detecting $\sim$1000 LAEs at $2.9<z<6.7$; (2) The deeper MUSE-Deep survey \citep{Bacon2017,Inami2017,Bacon2023} includes three tiers: a 10-hour $3\times3$~arcmin$^2$ mosaic (MOSAIC), a 31-hour $1\times1$~arcmin$^2$ region (UDF-10), and the 141-hour MUSE eXtremely Deep Field (MXDF; 1-arcmin diameter). Reduced with the \citet{Weilbacher2020} pipeline, our selection requires: for MUSE-Deep DR2 sources \citep{Bacon2023}, spectroscopic confidence ZCONF=2-3 (Ly$\alpha$ S/N$>5$-$7$) with validated \texttt{pyPlatefit} measurements; for MUSE-Wide \citep{Kerutt2022}, QtClassify confidence levels 2-3 (error rate $\lesssim10\%$) and HST-based Ly$\alpha$ equivalent width measurements. Overlapping sources prioritize \citet{Bacon2023} measurements.

For VANDELS \citep{Talia2023}, which provides $4800$--$9800$~\AA\ spectroscopy in CDFS and UDS fields, we select LAEs meeting: (1) Redshift confidence flags 3-4 ($>95\%$ reliability); (2) Ly$\alpha$ goodness-of-fit flag 1 (robust line measurements); (3) Positive Ly$\alpha$ fluxes. Ly$\alpha$ equivalent widths were derived via Gaussian profile fitting using \texttt{pylick} \citep{Borghi2022} following \citet{Kornei2010}.

The CANDELS-z7 sample \citep{Pentericci2018} completes our compilation, selected via identical photometric and spectroscopic criteria as the primary surveys. This multi-survey approach yields a comprehensive LAE sample with rigorously validated spectroscopic redshifts.

\begin{figure}[t!]
\centering
\includegraphics[width=\columnwidth]{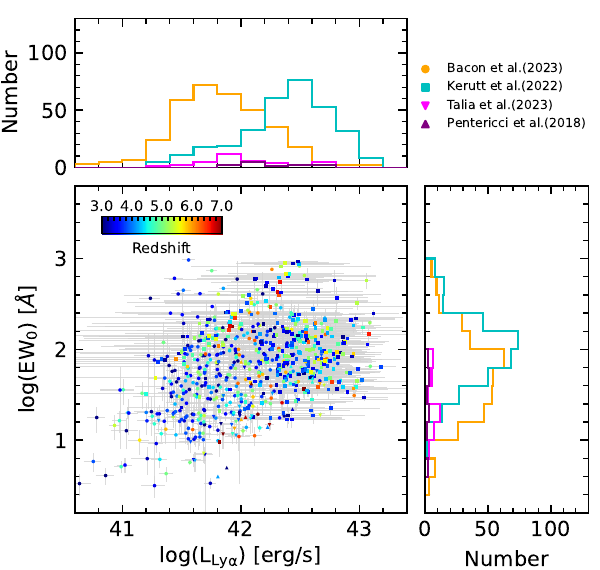}
\caption{The equivalent width versus Ly$\alpha$ luminosity for our sample of 817 LAEs.}
\label{FigVibStab}
\end{figure} 

\subsection{Sample Selection}

We began by compiling a large sample of spectroscopically confirmed LAEs from the GOODS-S field, 
selecting galaxies with secure Lyman-alpha emission-line detections ($S/N >3$) 
based on data from \citet{Pentericci2018, Kerutt2022, Bacon2023, Talia2023}. 
For sources common to both the MUSE-deep and MUSE-wide surveys, we prioritized 
the deeper MUSE-deep observations \citep{Bacon2023} to ensure higher spectral sensitivity and data consistency. 
The LAE candidates were then cross-matched with our JWST/NIRCam imaging data in the GOODS-S field. 
Detailed sample selection criteria are provided in Song et al. (2025, in preparation). 
To ensure robust multi-wavelength photometry, we retained only those sources detected 
in at least three NIRCam bands spanning F090W to F444W. This selection yielded a final sample of 817 LAEs.

The integrated fluxes of the LAEs in the HST bands (F435W, F606W, F814W, F105W, F125W, F140W, F160W) and JWST/NIRCam bands (F090W, F115W, F150W, F182M, F200W, F210M, F277W, F335M, F356W, F410M, F444W) were measured by integrating their radial surface brightness profiles. These profiles were computed using the Python package \texttt{photutils} \citep{Bradley2024} applied to PSF-matched images, with all contaminating sources masked to ensure accurate photometry.

\begin{figure*}[t!]
\centering
\includegraphics[width=\textwidth]{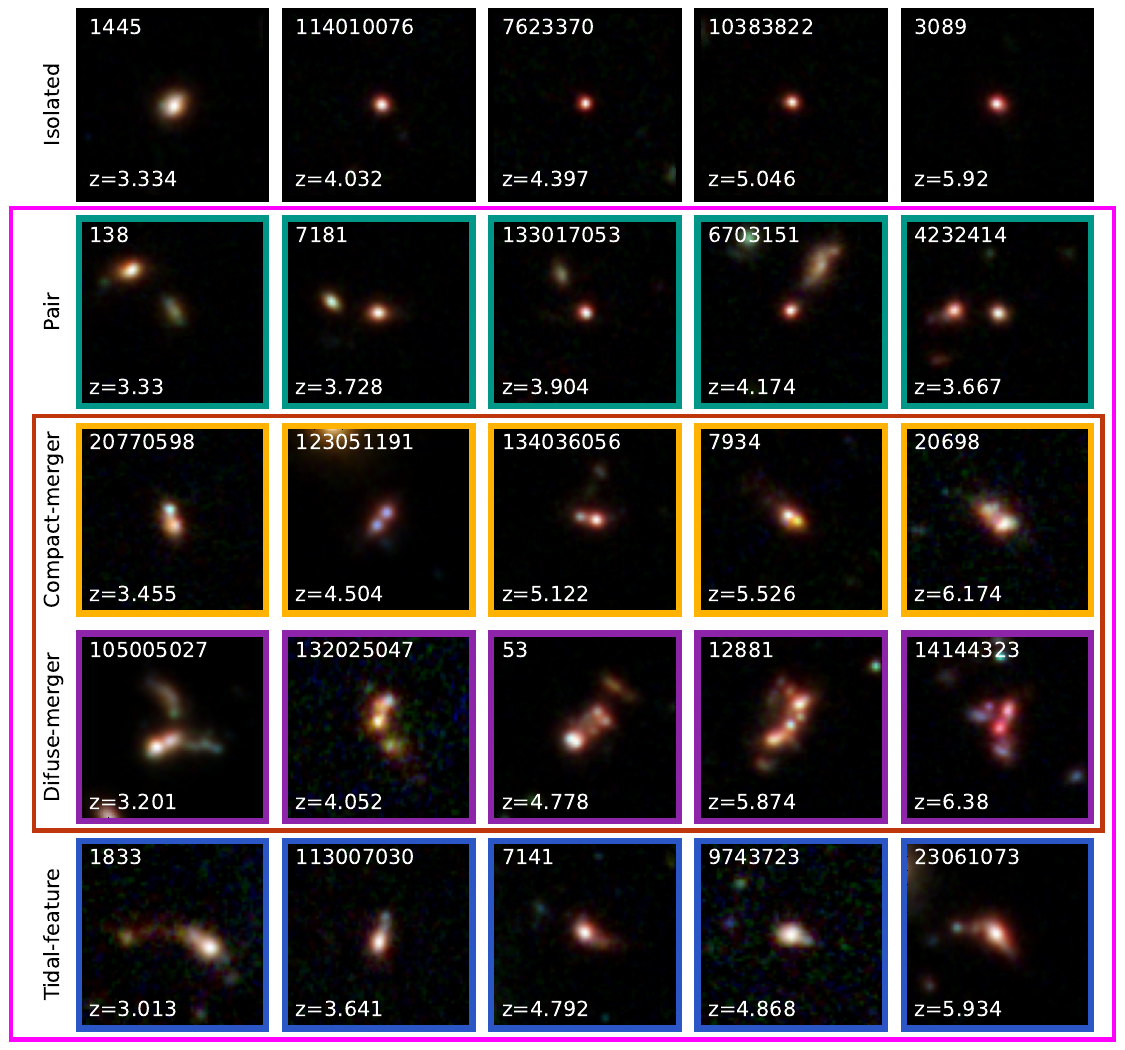} 
\caption{Example color images of our LAEs with visually classified morphologies. The field of view (FOV) 
is 2$\arcsec$$\times$2$\arcsec$. }
\label{sample_images}
\end{figure*}

\section{Methods} 
\label{sec:methods}
\subsection{SED Fitting}

The spectral energy distribution (SED) fitting for our LAE sample was performed using \texttt{CIGALE} (version 2022.0; \citet{Boquien2019,Yang2020,Yang2022}), incorporating photometric data spanning HST/F435W to JWST/NIRCam F444W.

We adopted the delayed-$\tau$ star formation history module (\texttt{sfhdelayed}) with parameter ranges constrained as follows: stellar population ages = 50-1500\,Myr, e-folding timescale $\tau$ = 30-5000\,Myr, and late burst mass fraction = 0.001-0.6. Stellar populations were modeled using \texttt{BC03} templates \citep{Bruzual2003} with sub-solar metallicities ($Z_{\star}/Z_\odot$ = 0.02-0.4) and a Chabrier initial mass function \citep{Chabrier2003}. Gas-phase properties were parameterized with metallicities $Z_{\rm gas}/Z_\odot$ = 0.02-0.6 and ionization parameters $\log U$ = $-3.5$ to $-1$, consistent with high-$z$ galaxy observations \citep{Maseda2020, Schaerer2022, Matthee2023, Brinchmann2023, Li2025}. Dust attenuation followed the modified Calzetti law \citep{Calzetti2000}, with color excess $E(B-V)$ = 0-0.6 scaled by a factor 0.44 to account for differential attenuation between stellar populations \citep{Charlot2000, Wild2011}. This configuration aligns with established CIGALE frameworks for LAE SED fitting \citep{Goovaerts2024a, Goovaerts2024b, Liu2024}.

\subsection{Non-parametric Morphological Measurements}

Non-parametric morphological indicators offer robust, quantitative diagnostics of galaxy structure. 
In this study, we focus on three merger-sensitive parameters: the Gini coefficient ($G$), the second-order moment of the brightest 20\% flux ($M_{20}$), and outer asymmetry ($A_{\rm O}$). $G$ quantifies flux concentration inequality, with higher values indicating more concentrated emission. $M_{20}$ traces spatial clustering of bright regions, sensitive to features like multiple nuclei. $A_{\rm O}$ specifically measures asymmetries in outer regions, efficiently identifying tidal features and diffuse merger remnants \citep{Lotz2004, Wen2014}.

Measurements follow established methodologies: the Gini coefficient ($G$) and $M_{20}$ were computed according to the prescriptions of \citet{Lotz2004}, while the outer asymmetry $A_{\rm O}$  was implemented 
following \citet{Wen2016} and \citet{Ren2023}. All morphological parameters were measured primarily in the F115W filter. For sources lacking reliable F115W data—due to non-coverage, contamination, or low signal-to-noise 
ratio (S/N $< 3$)—we adopted F090W for galaxies at $z<5$ and F150W for those at $z>5$. 
This wavelength-dependent substitution strategy ensures robust morphological characterization across the full redshift range while accounting for observational limitations.

\begin{figure*}[ht!]
\centering
\includegraphics[width=\textwidth]{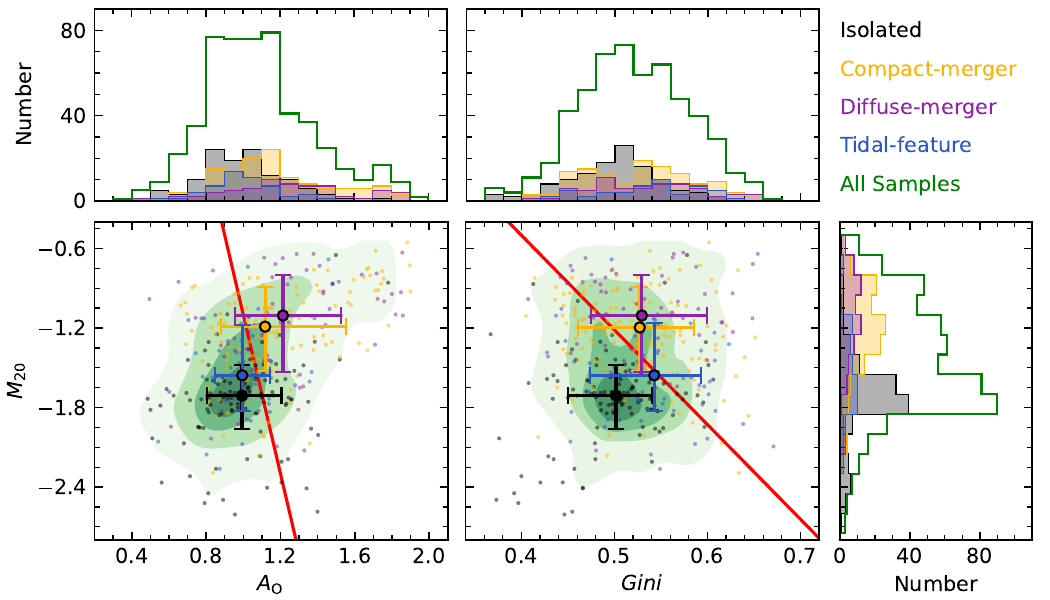}
\caption{
Diagnostic diagrams for securely classified Lyman-alpha emitters (LAEs) using non-parametric morphological indicators. 
\textbf{Left:} $M_{20}$ vs. $A_{\rm O}$ diagram, with the late-stage merger selection boundary (red line) defined by \citet{Ren2023}, incorporating redshift-dependent $A_{\rm O}$ corrections from \citet{Ren2024}.
\textbf{Right:} $M_{20}$ vs. Gini ($G$) diagram, adopting the merger criterion of \citet{Lotz2008} (red line). 
Large symbols with error bars represent the median values for each morphological class, with error bars indicating the 16th–84th percentile ranges. The analysis includes only morphologically secure sources with a signal-to-noise ratio SNR $> 3$. 
}
\label{Non-parametric}
\end{figure*} 

\subsection{Visual Morphological Classification}

We conducted visual morphological classification of LAEs using high-resolution JWST/NIRCam imaging spanning F090W to F444W. For each target, we generated $30 \times 30$ kpc$^2$ multi-band cutouts and constructed pseudo-color composites to enhance structural features. Classification followed these criteria:

\begin{itemize}
    \item \textbf{Isolated}: Single-galaxy systems within 10 kpc radius, or where neighboring galaxies have discordant redshifts;
    
    \item \textbf{Pair}: Galaxy pairs with projected separation $\leq 10$ kpc and consistent spectroscopic/photometric redshifts;
    
    \item \textbf{Compact-merger}: Systems exhibiting tightly bound, overlapping multiple components;
    
    \item \textbf{Diffuse-merger}: Extended interacting systems with resolved multiple galaxies and asymmetric profiles;
    
    \item \textbf{Tidal-feature}: Single-component galaxies displaying prominent tidal tails, shells, or disturbed features.
\end{itemize}

Two independent classifiers (JR and FL) performed the visual morphological assessments, with classifications 
deemed 'secure' only when both agreed unanimously. Discrepancies, occurring in less than 20\% of cases, 
were resolved through consensus review; any remaining unresolved cases were flagged as 'tentative' 
and excluded from statistical analyses. For evolutionary analysis, compact mergers and diffuse mergers 
were combined into a single \textit{late-stage merger} category to represent systems 
in advanced coalescence phases. Galaxies classified as pairs, late-stage mergers, or 
those exhibiting tidal features were collectively designated as \textit{interacting LAEs} 
for the purpose of interaction fraction analysis. Representative examples of each morphological class 
are presented in Figure~\ref{sample_images}.

\section{Results}
\label{sec:results}

\subsection{Nonparametric Morphological Analysis}
\label{ssec:nonparam}

\begin{figure*}[t!]
\centering
\includegraphics[width=\textwidth]{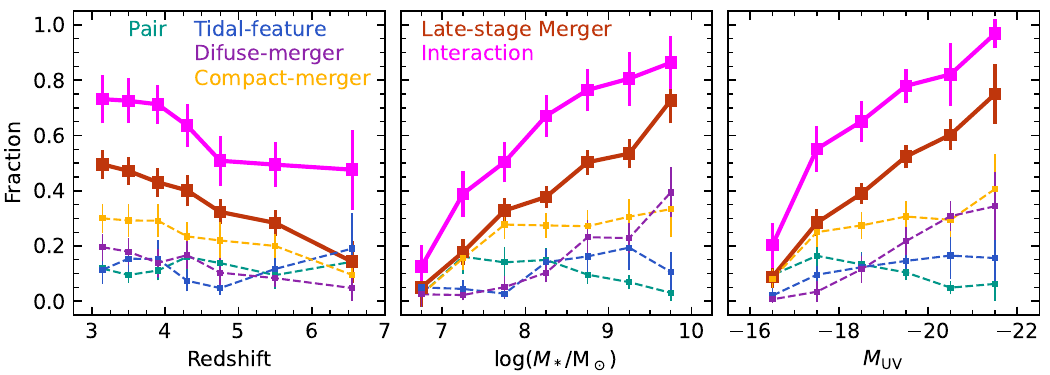}
\caption{
Merger fraction as functions of redshift ($z$), stellar mass ($M_*$), and UV magnitude ($M_{\rm UV}$). 
Late-stage mergers combine compact and diffuse mergers, while interacting systems include pairs, mergers, and tidal-feature galaxies. 
Error bars incorporate uncertainties from morphological identification and Poisson statistics.
}
\label{fig:merger_fractions}
\end{figure*} 

Non-parametric indicators--$M_{20}$ (second-order moment of the brightest 20\% flux), Gini coefficient ($G$), and outer asymmetry ($A_{\rm O}$)--provide quantitative diagnostics for merger identification. Based on visual classification, 
our LAEs are categorized into: isolated, compact-merger, diffuse-merger, and tidal-feature systems. Compact mergers with double nuclei are efficiently identified by the $G$-$M_{20}$ diagnostic \citep{Lotz2008}, while \citet{Ren2023}'s $M_{20}$-$A_{\rm O}$ plane captures both compact and diffuse late-stage mergers. 

Figure~\ref{Non-parametric} evaluates these criteria for our $z>3$ LAEs, considering only sources with image S/N $>3$ to minimize noise bias \citep{Lotz2006}. We note that while $G$ and $M_{20}$ show negligible redshift evolution, $A_{\rm O}$ increases by $\sim0.3$ from $z=0.5$ to $z=3$ \citep{Ren2024}. We therefore apply redshift-dependent corrections to the $A_{\rm O}$ threshold (red line, left panel) while retaining the original $G$-$M_{20}$ criterion (magenta line, right panel). 

Late-stage mergers predominantly occupy the selection regions with $M_{20}>-1.5$, $A_{\rm O}>1.1$, and $G>0.52$. Diffuse mergers show marginally higher $A_{\rm O}$ than compact mergers at comparable $M_{20}$, while tidal-feature systems share similar $G$ but lower $M_{20}$ values. Crucially, late-stage mergers separate cleanly from isolated systems in both diagrams, demonstrating that low-redshift merger criteria--despite being calibrated for massive galaxies--remain applicable to high-$z$, low-mass LAEs. This morphological invariance across cosmic time and mass scales suggests universal merger signatures (e.g., double nuclei, tidal features). The quantitative agreement between visual and parametric classifications further validates both methodologies.

\subsection{Merger Fraction Evolution}
\label{ssec:merger_frac}

Figure~\ref{fig:merger_fractions} reveals a strong evolution of late-stage merger and interaction fractions with redshift, stellar mass, and UV luminosity. 
The late-stage merger fraction increases from $\sim15\%$ at $z>6$ to 45-50\% at $z=3$, while the interaction fraction rises from $\sim45\%$ at $z>6$ to $70\%$ at $z<4$. 
The dependence on stellar mass is particularly pronounced: late-stage mergers rise from 
$\sim10\%$ at $\log(M_*/M_\odot)<7$ to 50\% at $\log(M_*/M_\odot)>9$, with interacting systems reaching 70-80\%. 
A similar trend is observed with UV luminosity: the late-stage merger fraction increases from 
$<10\%$ for faint LAEs ($M_{\rm UV}>-16$) to $70\%$ for luminous systems ($M_{\rm UV}<-21$), 
while the fraction of interacting LAEs approaches ubiquity ($>90\%$) at the bright end. 
For massive ($\log M_* > 8.5$) and luminous ($M_{\rm UV} < -19.5$) systems, 
merger fractions exceed 50\%, in agreement with previous HST/JWST studies \citep{Jiang2013,Zhu2024}. 
These universal trends establish mergers as dominant drivers in evolved, massive, luminous LAE populations.

\begin{figure*}[t!]
\centering
\includegraphics[width=\textwidth]{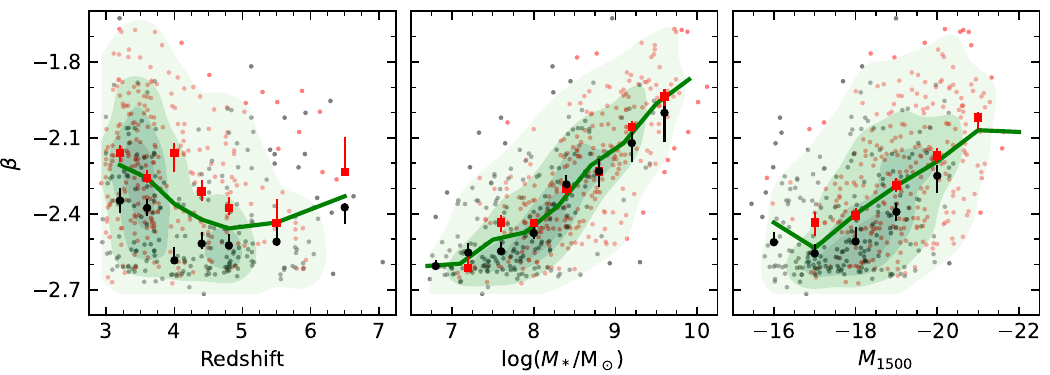}
\caption{
UV slope ($\beta$) versus redshift ($z$), stellar mass ($M_*$), and rest-frame 1500Å magnitude ($M_{\rm 1500}$). 
Black and red points denote isolated and late-stage merger LAEs, respectively, with green contours showing the full LAE distribution. 
Large symbols and lines represent median $\beta$ values in fixed bins: black (isolated), red (mergers), green (total sample).
}
\label{fig:UV_slope}
\end{figure*}

\subsection{UV Slopes}
\label{ssec:uv_slope}

Figure~\ref{fig:UV_slope} compares UV slopes ($\beta$) between isolated and late-stage merger LAEs. 
A general trend toward bluer slopes (lower $\beta$) is observed at higher redshifts ($3<z<7$). 
At fixed $z$, mergers exhibit marginally redder slopes than isolated systems ($\Delta\beta\sim0.2$), 
although this difference likely reflects intrinsic galaxy properties rather than direct merger-induced effects. 
When examined as a function of stellar mass (middle panels), both populations follow identical $\beta$-$M_*$ relations 
across $10^7$--$10^{10} M_\odot$, indicating that stellar mass—rather than merger status—governs UV spectral properties. 
t fixed $M_{\rm 1500}$ (right panels), isolated LAEs show slightly bluer slopes ($\Delta\beta\sim0.1$) than mergers. 
The apparent increase in $\beta$ at $z>5$ arises from the underrepresentation of low-mass LAEs in our high-$z$ sample. 
These results demonstrate that stellar mass is the primary determinant of UV slope, with mergers contributing only secondary modulation.

\subsection{Star formation properties}

\begin{figure*}[t!]
\centering
\includegraphics[width=0.45\textwidth]{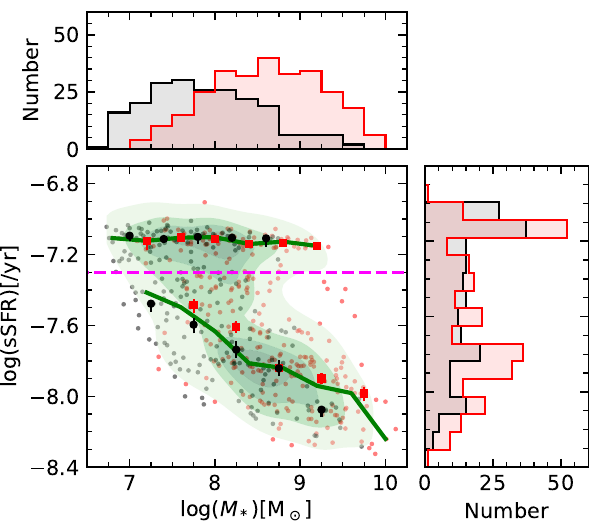} \qquad
\includegraphics[width=0.45\textwidth]{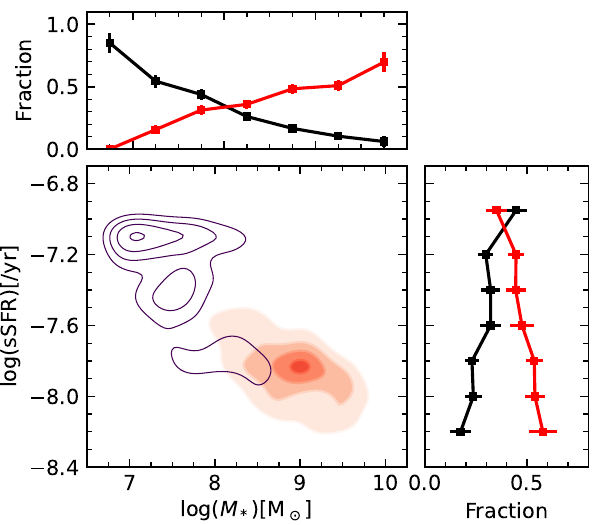}

\caption{
Left: Star Formation Main Sequence for isolated LAEs (black points), late-stage mergers (red points), and all LAEs (green contours). Black squares, red squares, and the green line represent median specific star formation rates in fixed stellar mass bins for isolated LAEs, late-stage mergers, and the full sample, respectively. Right: Two-dimensional distribution of the late-stage merger fraction (red colormap) and isolated LAE fraction (black contours) in the sSFR-$M_\star$ plane. 
}
\label{SFMS}
\end{figure*}

In the left panel of Figure~\ref{SFMS}, we present the relation between specific star formation rate (sSFR) and stellar mass ($M_\star$) for our LAE sample. Notably, both late-stage merging and isolated LAEs exhibit similar median sSFR values within the same stellar mass bins, indicating no significant enhancement of star formation activity in late-stage mergers relative to isolated LAEs.

Intriguingly, the sSFR-$M_\star$ distribution of our full LAE sample reveals a distinct bimodal structure, consistent with results reported by \cite{Iani2024}. This bimodality reflects two dominant populations: relatively young and old LAEs. Young LAEs are characterized by lower stellar masses and higher sSFRs, whereas old LAEs exhibit the opposite trend—higher masses and lower sSFRs. Specifically, our isolated LAEs have a median stellar mass of $\log(M_\star/M_\odot)=7.8$ and a median sSFR of $\log(\mathrm{sSFR}/\mathrm{yr}^{-1})=-7.4$, while  while merging LAEs are more massive, with $\log(M_\star/M_\odot)=8.6$, and slightly lower in sSFR, $\log(\mathrm{sSFR}/\mathrm{yr}^{-1})=-7.6$. The right panel of Figure~\ref{SFMS} illustrates the spatial distribution of isolated and late-stage merging LAEs in the sSFR--$M_\star$ plane. Notably, late-stage merging LAEs dominate over isolated systems at $\log(M_\star/M_\odot)>8.0$, indicating a preferential association with massive, evolved galaxies. In contrast, compact isolated LAEs predominantly represent low-mass, actively star-forming systems. Together, these findings reveal a clear evolutionary dichotomy within the LAE population.

\section{Discussions} \label{sec:discussion}

\subsection{Impact of Cosmic Dimming on Merger Identification}

As redshift increases, galaxy surface brightness declines as $\propto (1+z)^4$, rendering low-surface-brightness tidal features increasingly difficult to detect. This may lead to an underestimation of tidal-feature Lyman-$\alpha$ emitters (LAEs) at high redshifts. However, tidal features correspond to transient merger phases and have intrinsically low occurrence rates--only $\sim$10\% in our sample. Consequently, even accounting for observational incompleteness, the true fraction of such systems satisfies $f_{\text{true}} < 0.10$. Furthermore, tidal-feature LAEs represent only a subset of all interacting LAEs and thus contribute modestly to the overall evolution of merger fractions. Crucially, while observational biases against tidal-feature detection would imply a \textit{higher} true interaction rate, this correction does not materially affect our primary conclusions.

Our conclusions are primarily based on a rigorously selected sample of late-stage merger LAEs. These systems are identified by the presence of two or more compact nuclei or extended, multi-component structures—morphological features that are relatively resilient to surface brightness dimming. As a result, the observational limitations discussed above have negligible impact on our results. The robustness of the late-stage merger sample further strengthens the reliability of our findings regarding the evolutionary trends of merger fractions in LAE populations.

\subsection{Nonparametric morphological indentifaication of Mergers}

The measurement of non-parametric morphological parameters (e.g., Gini, \( M_{20} \), $A_{\rm O}$) 
is inherently sensitive to image resolution (quantified by the PSF FWHM) and signal-to-noise ratio (S/N) \citep{Lotz2006, Ren2024}. In our analysis, these parameters are measured from JWST/NIRCam F115W imaging with a PSF FWHM 
of $ 0.\prime\prime08$, which is comparable to or slightly better than the resolution of HST/F814W imaging ($ 0.\prime\prime09$)—-the dataset used to establish the original Gini-\( M_{20} \) and$A_{\rm O}$--\( M_{20} \) merger selection criteria \citep{Lotz2004, Lotz2008, Ren2023}. Additionally, we impose a stringent $S/N>3$ threshold to minimize noise-induced biases in parameter measurements. These considerations ensure that our morphological measurements are robust against instrumental and observational artifacts.

The efficacy of merger identification using non-parametric criteria depends critically on their sensitivity to different interaction phases. For example, \( M_{20} \) is particularly effective at identifying mergers with multicomponent structures (e.g., double nuclei), whereas $A_{\rm O}$ is more sensitive to extended tidal features \citep{Lotz2004, Wen2014}. This explains why our late-stage mergers—selected based on visual signatures of multiple components—are robustly identified by \( M_{20} \)-based selection criteria. Notably, diffuse mergers exhibiting prominent tidal tails show higher $A_{\rm O}$ values compared to compact systems, consistent with their more advanced dynamical states. The partial overlap with \citet{Zhu2024}'s LyC leaker sample—where only 7 out of 21 visually classified mergers are detected via the Gini--$M_{20}$ criteria—likely reflects differences in morphological selection priorities. Their sample includes systems with faint tidal features but lacking clear multicomponent structure, which are less efficiently captured by classical Gini--$M_{20}$ diagnostics. Importantly, these tidal-feature-dominated mergers exhibit distinct Gini distributions compared to isolated LAEs, suggesting that revised Gini--$M_{20}$ thresholds could enhance the completeness of merger identification in such cases.

Remarkably, our high-$z$ LAE mergers satisfy the same Gini--$M_{20}$/$A_{\rm O}$--$M_{20}$ criteria established for massive galaxies at $z<1$, indicating morphological homology in gas-rich mergers across cosmic time. 
This consistency validates the use of nonparametric morphological indicators for identifying mergers up to $z \sim 7$, provided that parameters--such as Gini and $M_{20}$--are chosen for their stability against surface-brightness dimming \citep{Ren2024}. Given redshift-dependent variations in asymmetry ($\delta A_{\rm O}/\delta z > 0$) and clumpiness \citep{Rose2023}, the careful selection of appropriate morphological indicators remains essential for reliably identifying diverse merger populations across different cosmic epochs.

\subsection{The role of mergers in the formation of LAEs}

In the $\mathrm{sSFR}-M_*$ plane, LAEs exhibit a bimodal distribution that has been robustly established through observations \citep{Iani2024}. This bimodality separates LAEs into two distinct populations—low-mass, high-sSFR systems ($\mathrm{sSFR} > 10^{-8}$ yr$^{-1}$) indicative of young stellar populations, and massive, low-sSFR galaxies ($\mathrm{sSFR} < 10^{-9}$ yr$^{-1}$) characteristic of evolved systems—a distinction consistently observed in both observational \citep{Lai2008,Finkelstein2009,Pentericci2009,Nilsson2009,Iani2024} and simulation studies \citep{Shimizu2010}. Crucially, our morphological analysis reveals that this dichotomy extends beyond star formation activity to structural properties: low-mass LAEs are predominantly isolated, while approximately $\sim$50\% of massive LAEs ($\log(M_*/M_\odot) > 8.5$) show clear signatures of late-stage mergers. 

To investigate potential correlations between stellar age and morphology, we classifed LAEs into two populations following \citet{Shimizu2010}: Type~1 (young LAEs; $\mathrm{Age} < 150\,\mathrm{Myr}$) and Type~2 (old LAEs; $\mathrm{Age} \geq 150\,\mathrm{Myr}$). We performed Kolmogorov-Smirnov (K-S) tests on the distributions of stellar mass ($\log(M_*)$) and specific star formation rate ($\log(\mathrm{SSFR})$) across these categories. The resulting p-values are listed in Table \ref{K-S}. For stellar mass, the K-S test comparing old LAEs and mergers yields a p-value of 0.16, indicating no statistically significant difference in their stellar mass distributions ($p > 0.05$). This suggests that old LAEs and mergers share similar stellar mass distributions, consistent with the hypothesis that mergers are preferentially found in more massive systems, irrespective of stellar age. In contrast, the comparison of $\log(\mathrm{SSFR})$ between young LAEs and isolated systems yields a p-value of 0.31, also failing to reject the null hypothesis ($p > 0.05$). This implies that isolated LAEs exhibit star formation activity statistically indistinguishable from that of young LAEs, supporting a scenario in which isolated systems may retain youthful star formation histories despite the absence of recent interactions. Collectively, these results suggest a nuanced relationship: stellar mass is more closely linked to merger signatures among old LAEs, while isolated LAEs resemble young populations in terms of sSFR. This bimodality reflects a complex interplay between stellar age, morphological structure, and evolutionary pathways.

The redshift evolution further supports this paradigm. The observed increase in merger fraction from 15\% at $z > 6$ to 45\% at $z = 3$ (Figure 5) closely mirrors simulations predicting the emergence of Type 2 LAEs--massive, merger-driven systems--as dark matter halos grow through hierarchical assembly \citep{Shimizu2010}. The strong agreement between our observational results and theoretical models underscores mergers as the primary driver of 
Ly$\alpha$ detectability in massive galaxies. This convergence establishes a unified framework in which galaxy interactions shape both the demographic distribution and spectral properties of LAEs across cosmic time. In contrast, young, low-mass, and predominantly isolated LAEs dominate at higher redshifts. In summary, our findings suggest that low- and high-redshift LAEs represent fundamentally distinct galaxy populations: high-redshift LAEs are typically young, low-mass, and primordial systems, whereas their low-redshift counterparts are more likely to be evolved, massive galaxies with older stellar populations shaped by mergers.

The exceptionally high merger fraction observed in massive LAEs ($f_{\mathrm{merger}} \sim 80\%$) stands in stark contrast to the $\sim20\%$ interaction rate reported in coeval non-LAE galaxies \citep{Duan2024}. 
When accounting for typical merger visibility timescales, our results suggest that nearly all massive LAEs undergo at least one significant interaction during their evolution. These merger-driven processes play a critical role in enhancing both Ly$\alpha$ production--via triggered star formation--and Ly$\alpha$ escape, through multiple physical channels such as gas dispersal \citep{Witten2024}, regulation of outflows, and enhanced leakage of Lyman-continuum (LyC) photons \citep{Zhu2024,Kostyuk2024}. Supporting this scenario, \citet{Shimizu2025} demonstrated elevated Ly$\alpha$ escape efficiency specifically in old LAEs—-precisely the population we identify as predominantly merger-driven. This convergence of observational and theoretical evidence firmly establishes galaxy mergers as pivotal regulators of Ly$\alpha$ escape physics, effectively transforming otherwise ordinary star-forming galaxies into detectable LAEs by simultaneously boosting Ly$\alpha$ production and clearing escape pathways.

\begin{table}[t!]
\centering
\scriptsize
\caption{ The p-value of K-S test.}
\label{K-S}
\begin{tabular}{|c|c|c|}
\hline 
 Types   & Isolated &  Late-stege Merger \\
\hline

Young  & p(${M_*}$)=0\,\,\,P(sSFR)=0.31  & p(${M_*}$)=0\,\,\,P(sSFR)=0.0  \\

\hline

Old   & p(${M_*}$)=0\,\,\,P(sSFR)=0  & p(${M_*}$)=0.16\,\,\,P(sSFR)=0.0  \\

\hline

\end{tabular}
 \end{table}

\section{Summary}  \label{sec:conclusions}
Utilizing \textit{JWST}/NIRCam multi-band imaging, we present a systematic analysis of merger activity in 817 spectroscopically confirmed Lyman-$\alpha$ emitters (LAEs) at $3 < z < 7$ within the GOODS-S field. By combining spectral energy distribution (SED) fitting with visual classification of morphological signatures of mergers, we identify three key results:

\begin{itemize}

\item Late-stage mergers exhibit distinct morphological signatures in \textit{JWST}/NIRCam F115W imaging 
characterized by the following nonparametric criteria: $M_{20} > -1.4$ (indicative of multi-component structures), $A_{\rm O} > 1.0$ (reflecting high asymmetry), and Gini $> 0.52$ (signifying enhanced light concentration). 
These quantitative measures confirm that approximately $\sim$80\% of visually classified mergers lie within the predicted region of merger parameter space.

\item The late-stage merger fraction is 39.4\% $\pm$ 2.5\%, and the interaction fraction reaches up to 
60.6\% $\pm$ 6.3\%. Both fractions exhibit strong evolution with redshift ($z$), stellar mass ($M_*$), and UV magnitude ($M_{\rm UV}$). For systems with $\log M_* > 8.5$ and $M_{\rm UV} < -19.5$, merger fractions exceed 50\%. 
At the highest stellar masses and UV luminosities, the interaction fraction can reach $\sim$80\%. 

\item  Isolated and merging LAEs exhibit minimal differences in UV slope ($\beta$) when matched in $M_*$ and $M_{\rm UV}$, suggesting that mergers do not fundamentally alter the properties of our sample. All LAEs in the sample display a bimodal distribution in $\mathrm{sSFR}-M_*$ plane. 
\end{itemize}

We propose that low-mass isolated LAEs represent the early stages of galaxy assembly, whereas 
massive LAEs likely originate from typical SFGs undergoing mergers that temporarily enhance Ly$\alpha$ photon escape efficiency. These results underscore mergers as critical drivers of both LAE visibility and stellar mass 
assembly in the early universe. Nevertheless, their influence on intrinsic properties remains secondary to 
the underlying scaling relations governed by stellar mass ($M_*$) and UV luminosity ($M_{\text{UV}}$). 

\section*{acknowledgements}

The data were obtained from the Mikulski Archive for Space Telescopes at the Space Telescope Science Institute, which is operated by the Association of Universities for Research in Astronomy, Inc., under NASA contract NAS 5-03127 for JWST.  This paper makes use of the following JWST data:  Programs 1176, 1180, 1210, 1283, 1286, 1895, 1963, 2079, 2198, 2514, 2516, 3215, 3990, and 654. NL acknowledges the support from the Ministry of Science and Technology of China (No. 2020SKA0110100), the science research grants from the China Manned Space Project (No. CMS-CSST-2021-A01), and the CAS Project for Young Scientists in Basic Research (No. YSBR-062). This project is supported by the National Natural Science Foundation of China (NSFC grants No. 12273052, 11733006,12090040, 12090041, and 12073051) and the science research grants from the China Manned Space Project (No. CMS-CSST-2021-A04). 

\bibliography{LAE}{}
\bibliographystyle{aasjournal}

\end{document}